\documentclass[10pt,a4paper]{article}
\usepackage[dvips]{color}
\usepackage{epsfig}
\usepackage{amsmath}
\usepackage{graphicx}
\usepackage{authblk}
\usepackage{amssymb,amsmath}
\usepackage{amsmath}

\begin{document}
\title{\bf Quintessence Cosmology with an Effective $\Lambda$-Term in Lyra Manifold}
\author{{M. Khurshudyan$^{a}$ \thanks{Email:
khurshudyan@yandex.ru}, \hspace{1mm} A. Pasqua$^{b}$ \thanks{Email: toto.pasqua@gmail.com}, \hspace{1mm} J. Sadeghi$^{c}$ \thanks{Email: pouriya@ipm.ir},\hspace{1mm}
and H. Farahani$^{c}$ \thanks{Email:
h.farahani@umz.ac.ir}}\\
$^{a}${\small {\em Department of Theoretical Physics, Yerevan State
University, 1 Alex Manookian, 0025, Yerevan, Armenia}}\\
$^{b}${\small {\em Department of Physics, University of Trieste, Via Valerio, 2 34127 Trieste, Italy}}\\
$^{c}${\small {\em Department of Physics, Mazandaran University, Babolsar, Iran}}\\
{\small {\em P .O .Box 47416-95447, Babolsar, Iran}}\\ }  \maketitle
\begin{abstract}
In this paper, we study quintessence cosmology with an effective $\Lambda$-term in Lyra manifold. We consider three different models by choosing variable $\Lambda$ depend on time, the Hubble parameter and the energy density of dark matter and dark energy. Dark energy assumed as quintessence which interacts with the dark matter. By using numerical analysis we investigate behavior of cosmological parameters in three different models and compare our results with observational data. Statefinder diagnostic is also performed for all models.
\end{abstract}

\section{\large{Introduction}}
Accelerated expansion of universe can described by dark energy which has positive energy and negative pressure [1, 2]. There are several theories to describe the dark energy such as Einstein's
cosmological constant which has two crucial problems so called fine tuning and
coincidence [3]. There are also
other interesting models to describe the dark energy such as
$k$-essence model [4], tachyonic models [5] and Chaplygin gas models [6-20]. An interesting model to describe dark energy is called quintessence [21-23]. Quintessence is described by a canonical scalar field $\phi$ minimally coupled to gravity.
Compared to other scalar-field models such as phantoms and k-essence, quintessence
is the simplest scalar-field scenario without having theoretical problems such as the
appearance of ghosts and Laplacian instabilities.\\
On the other hand, The Lyra's geometry provides one of the possible
alternatives in modification of the cosmological models [24]. Such a modification of the gravitational theory has long
been known, but now it again attracts attention due
to the opening of the late-time cosmological acceleration. The effective cosmological term in Lyra's
geometry recently has been studied by the several works, such as [25, 26].\\
Moreover, it is well known that Einstein equations of general relativity do
not permit any variations in the gravitational constant $G$ and
cosmological constant $\Lambda$ because of the fact that the
Einstein tensor has zero divergence and energy conservation law is
also zero. So, some modifications of Einstein equations are
necessary. Therefore, the study of the varying $G$ and $\Lambda$ can
be done only through modified field equations and modified
conservation laws. Already we construct several cosmological models based on variation of $G$ and $\Lambda$ [27-31].\\
Finally we should say about importance of interactions in cosmological models. One of the ways to solve the cosmological coincidence
problem is to consider the interaction between the components on phenomenological level. Also by consideration of interaction between dark matter and dark energy we can construct a real model of universe [32-35].\\
This paper is organized as the follows. In section 2 we review quintessence cosmology, and in section 3 we introduce our models. In section 4 we write field equations which should solved to obtain behavior of cosmological parameters. In section 5 we consider special case of constant $G$ and $\Lambda$. In section 6 we give results of our numerical analysis about cosmological parameters in three different models. In section 7 statefinder diagnostic will perform and numerically will analyze. Finally, in section 8 we give conclusions.

\section{\large{Quintessence cosmology}}
Quintessence is a scalar field model for the dark energy described by the field $\phi$ and the potential $V(\phi)$. Energy density and pressure are given as the follow,
\begin{equation}\label{eq:rhoQ}
\rho_{Q}=\frac{1}{2}\dot{\phi}^{2}+V(\phi),
\end{equation}
and,
\begin{equation}\label{eq:rhoP}
\rho_{b}=\frac{1}{2}\dot{\phi}^{2}-V(\phi),
\end{equation}
where $\rho_{Q}$ denotes the  density of dark energy and $\rho_{b}$ denotes the density of a barotropic fluid which will described dark matter in universe with the equation of state of the form $P_{b}=\omega_{b}\rho_{b}$. We consider a model for the universe where an effective energy density and pressure assumed to be given as the follows,
\begin{equation}\label{eq:rhoeff}
\rho=\rho_{Q}+\rho_{b},
\end{equation}
and,
\begin{equation}\label{eq:Peff}
P=P_{Q}+P_{b}.
\end{equation}

\section{\large{The models}}
First of all we introduce an interaction $Q$ between dark energy and dark matter as follow,
\begin{equation}\label{eq:Q}
Q=3Hb(\rho_{b}+\rho_{Q}),
\end{equation}
where $b$ is a positive constant. The solving strategy and structure of the problem is the following that we will assume that the form of the potential $V(\phi)$ is given,
\begin{equation}\label{eq:potV}
V(\phi)=V_{0}e^{ \left [-\frac{\alpha}{2}\phi^{\gamma} \right ]},
\end{equation}
where $\alpha$ and $\gamma$ are arbitrary constants.
Moreover, we will consider 3 different forms of $\Lambda$ as the follows. In the first model we assume that,
\begin{equation}\label{7}
\Lambda=\rho_{Q}+\rho_{b}e^{[-tH]}.
\end{equation}
It is clear that at the late time, where the density of dark energy is infinitesimal constant, the value of $\Lambda$ also takes infinitesimal constant, which may be agree with observational data.\\
In the second model we assume that,
\begin{equation}\label{8}
\Lambda=H^{2}+(\rho_{b}+\rho_{Q})e^{[-tH]}.
\end{equation}
Again, after late time, the value of $\Lambda$ takes the value $H_{0}^{2}$. Finally in the third model we consider the following relation,
\begin{equation}\label{9}
\Lambda=t^{-2}+\rho_{Q}+\rho_{b}e^{[-tH]}.
\end{equation}
The late time behavior of this model is similar to the first model.\\
According to the relations (7), (8) and (9) we have three different models which will analyze in the next sections.
\section{\large{The field equations}}
Field equations that govern our model of consideration are,
\begin{equation}\label{eq:Einstein eq}
R_{\mu\nu}-\frac{1}{2}g_{\mu\nu}R-\Lambda g_{\mu \nu}+\frac{3}{2}\phi_{\mu}\phi_{\nu}-\frac{3}{4}g_{\mu \nu}\phi^{\alpha}\phi_{\alpha}=T_{\mu\nu}.
\end{equation}
Considering the content of the universe to be a perfect fluid, we have,
\begin{equation}\label{eq:T}
T_{\mu\nu}=(\rho+P)u_{\mu}u_{\nu}-Pg_{\mu \nu},
\end{equation}
where $u_{\mu}=(1,0,0,0)$ is a 4-velocity of the co-moving
observer, satisfying $u_{\mu}u^{\mu}=1$. Let $\phi_{\mu}$ be a time-like
vector field of displacement,
\begin{equation}
\phi_{\mu}=\left ( \frac{2}{\sqrt{3}}\beta,0,0,0 \right ),
\end{equation}
where $\beta=\beta(t)$ is a function of time alone, and the factor $\frac{2}{\sqrt{3}}$ is substituted in order to simplify the writing of all the following equations.
By using FRW metric for a flat universe,
\begin{equation}\label{s2}
ds^2=-dt^2+a(t)^2\left(dr^{2}+r^{2}d\Omega^{2}\right),
\end{equation}
the field equations can be reduced to the following Friedmann equations,
\begin{equation}\label{eq:f1}
3H^{2}-\beta^{2}=\rho+\Lambda,
\end{equation}
and,
\begin{equation}\label{eq:Freidmann2}
2\dot{H}+3H^{2}+\beta^{2}=-P+\Lambda,
\end{equation}
where $H=\frac{\dot{a}}{a}$ is the Hubble parameter, and dot
stands for differentiation with respect to cosmic
time $t$, $d\Omega^{2}=d\theta^{2}+\sin^{2}\theta d\phi^{2}$, and $a(t)$
represents the scale factor. The $\theta$ and $\phi$ parameters are
the usual azimuthal and polar angles of spherical coordinates, with
$0\leq\theta\leq\pi$ and $0\leq\phi<2\pi$.\\
The continuity equation reads as,
\begin{equation}\label{eq:coneq}
\dot{\rho}+\dot{\Lambda}+2\beta\dot{\beta}+3H(\rho+P+2\beta^{2})=0.
\end{equation}
If we assume that,
\begin{equation}\label{eq:DEDM}
\dot{\rho}+3H(\rho+P)=0,
\end{equation}
then, Eq. (\ref{eq:coneq}) will give a link between $\Lambda$ and $\beta$ of the following form,
\begin{equation}\label{eq:lbeta}
\dot{\Lambda}+2\beta\dot{\beta}+6H\beta^{2}=0.
\end{equation}
To introduce an interaction between the dark energy and dark matter, we should mathematically split Eq. (\ref{eq:DEDM}) into two following equations,
\begin{equation}\label{eq:inteqm}
\dot{\rho}_{b}+3H(\rho_{b}+P_{b})=Q,
\end{equation}
and,
\begin{equation}\label{eq:inteqG}
\dot{\rho}_{Q}+3H(\rho_{Q}+P_{Q})=-Q.
\end{equation}
For the barotropic fluid with $P_{b}=\omega_{b}\rho_{b}$, equation (\ref{eq:inteqm}) will take the following form,
\begin{equation}
\dot{\rho}_{b}+3H(1+\omega_{b}-b)\rho_{b}=3Hb\rho_{Q}.
\end{equation}
Cosmological parameters of our interest are EoS parameters of each fluid components $\omega_{i}=P_{i}/\rho_{i}$, EoS parameter of composed fluid,
\begin{equation}
\omega_{tot}=\frac{P_{b}+P_{Q} }{\rho_{b}+\rho_{Q}},
\end{equation}
deceleration parameter $q$, which can be written as,
\begin{equation}\label{eq:accchange}
q=\frac{1}{2}(1+3\frac{P}{\rho} ).
\end{equation}

\section{\large{Case of constant $G$ and $\Lambda$}}
Before analyzing our three models we assume the simplest case with constant $G$ and $\Lambda$ to obtain effect of varying $\Lambda$ in the next section.
According to this assumption, Eq.(\ref{eq:coneq}) will be modified as the following,
\begin{equation}
\dot{\rho}+2\beta\dot{\beta}+3H(\rho+P+2\beta^{2})=0,
\end{equation}
with $\dot{\rho}+3H(\rho+P)=0$, we have,
\begin{equation}
2\dot{\beta}+6H\beta=0.
\end{equation}
Numerical analysis of the model gives the following behavior for the model.\\
From the Fig. 1 we can see that the Hubble expansion parameter is decreasing function of time and yields to a constant at the late time. It is illustrated that the interaction term increases value of the Hubble expansion parameter. Fig. 1 also contains deceleration parameter. We can see acceleration to deceleration phase transition. In the case of non-interacting case ($b=0$) the final value of of deceleration parameter is positive which disagree with current observations. Therefore, presence of interaction is necessary to obtain $q\rightarrow-1$ or $q\geq-1$ in agreement with observational data.\\
Behavior of EoS parameters plotted in the Fig. 2. In this case also we find that presence of interaction term is necessary to obtain $\omega\rightarrow-1$. In the case of non-interacting component we see that $\omega\rightarrow0$.\\
Finally in the plots of the Fig. 3 we see behavior of the scalar field and the potential. We see that scalar field in increasing function of time and decreased it's value by increasing interaction strength. The situation for the potential in inverse. It is decreasing function of time which vanishes at the late time, and it's value increased by increasing interaction strength.\\
We will compare our next results with the results of this section to obtain behavior of varying $\Lambda$ in the models.

\begin{figure}[h!]
 \begin{center}$
 \begin{array}{cccc}
\includegraphics[width=50 mm]{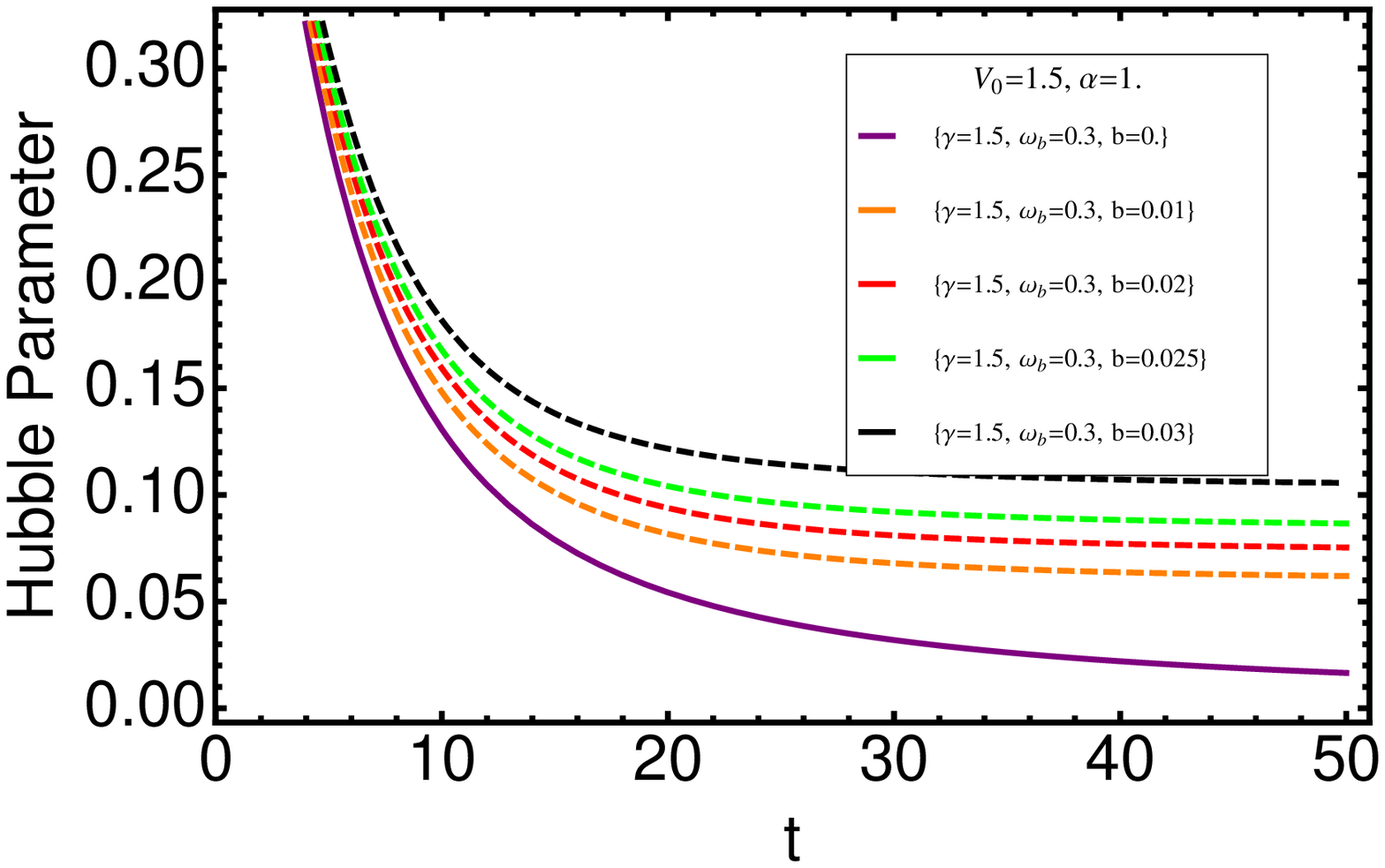} &
\includegraphics[width=50 mm]{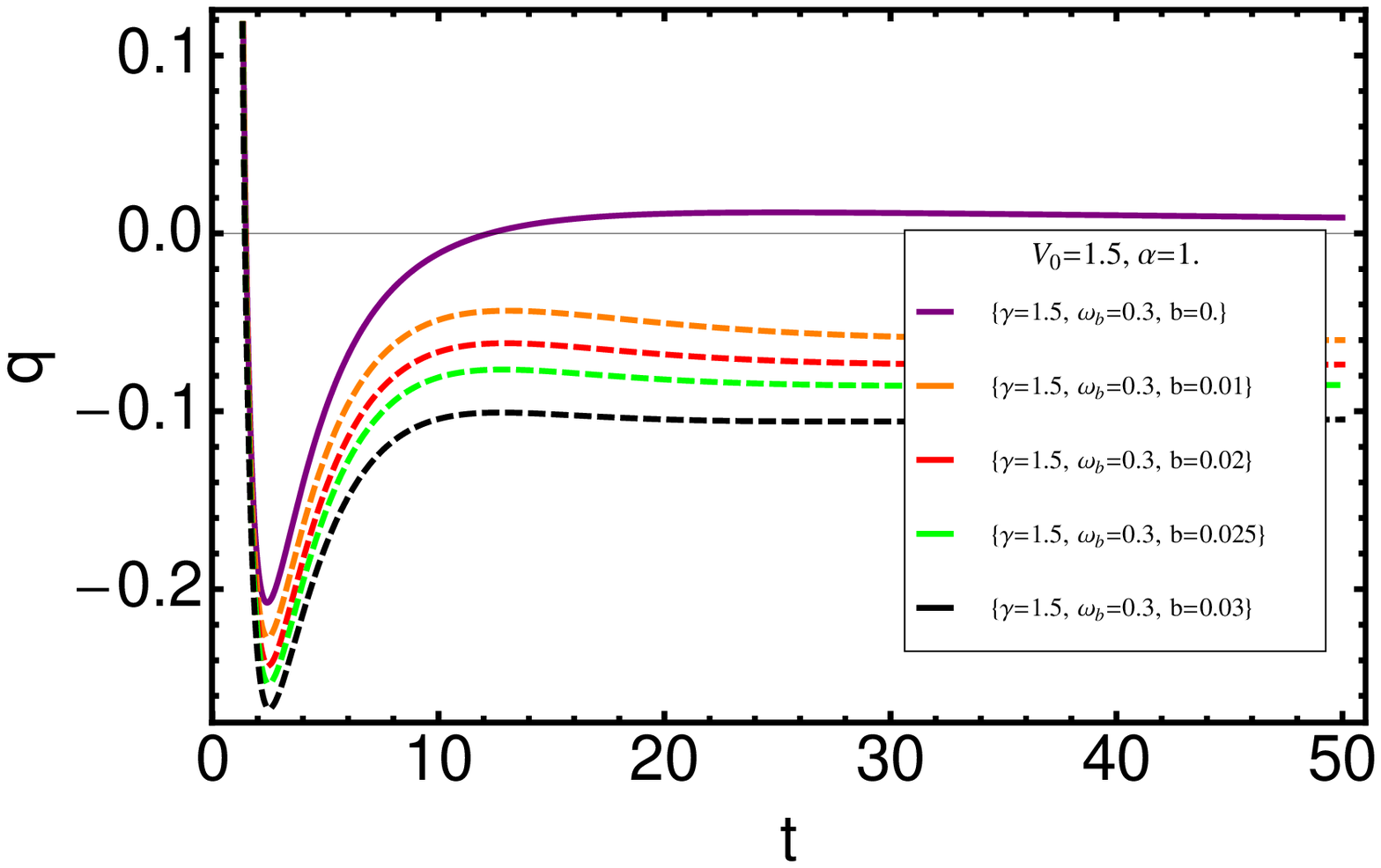}
 \end{array}$
 \end{center}
\caption{Behavior of Hubble parameter $H$ and deceleration parameter $q$ against $t$ for constant $G$ and $\Lambda$.}
 \label{fig:const1}
\end{figure}

\begin{figure}[h!]
 \begin{center}$
 \begin{array}{cccc}
\includegraphics[width=50 mm]{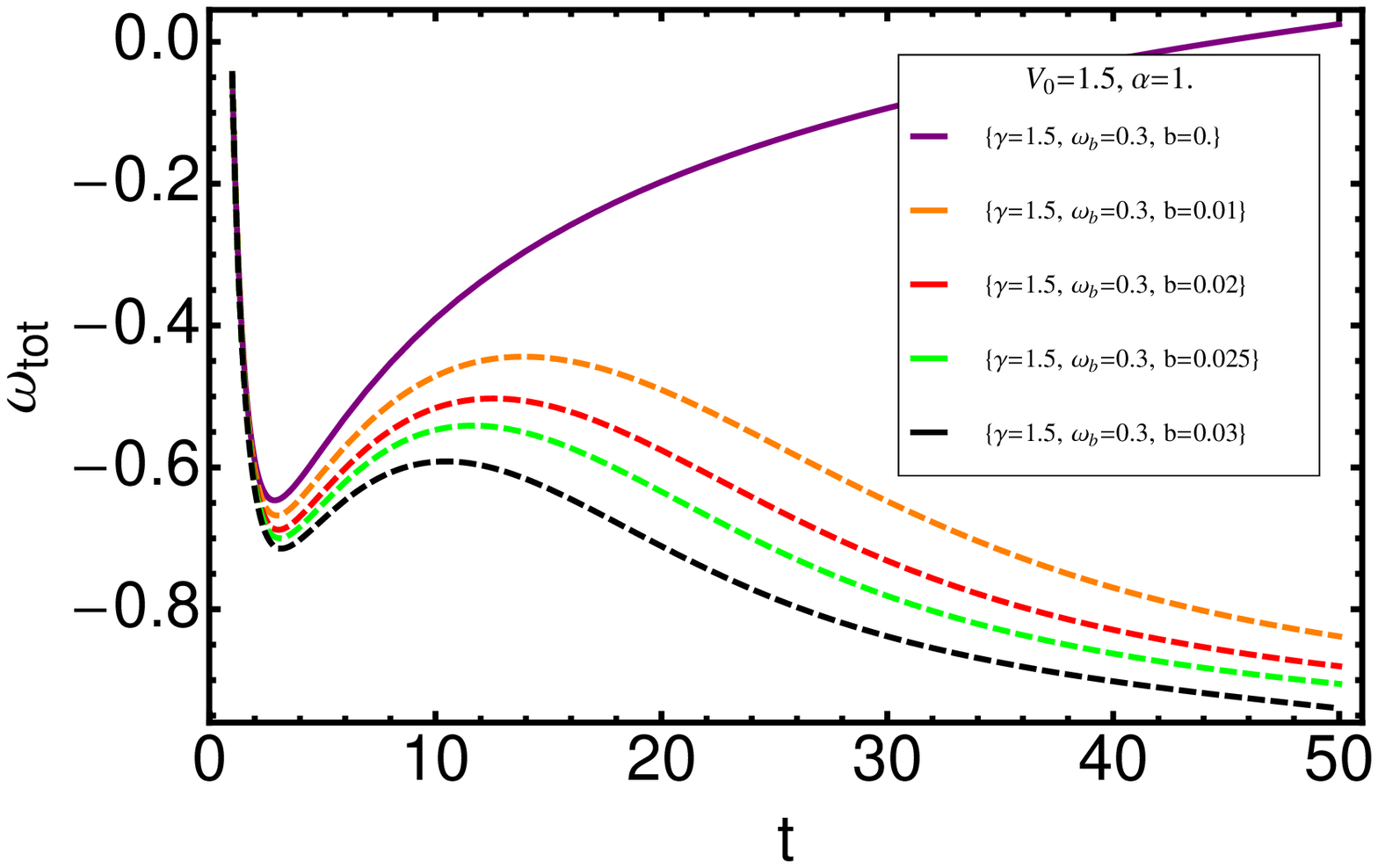} &
\includegraphics[width=50 mm]{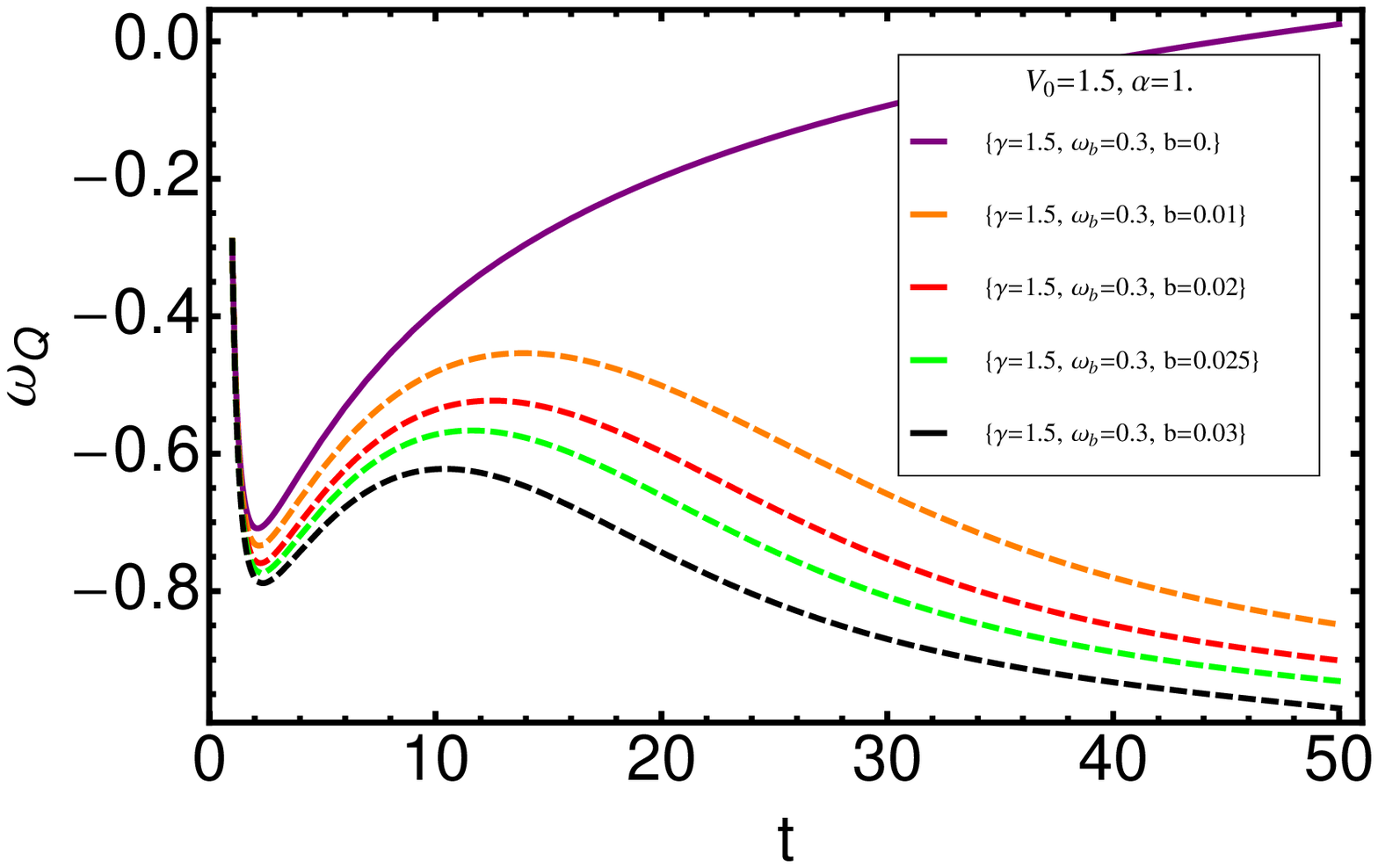}
 \end{array}$
 \end{center}
\caption{Behavior of EoS parameter $\omega_{tot}$ and $\omega_{Q}$ against $t$ for constant $G$ and $\Lambda$.}
 \label{fig:const2}
\end{figure}

\begin{figure}[h!]
 \begin{center}$
 \begin{array}{cccc}
\includegraphics[width=50 mm]{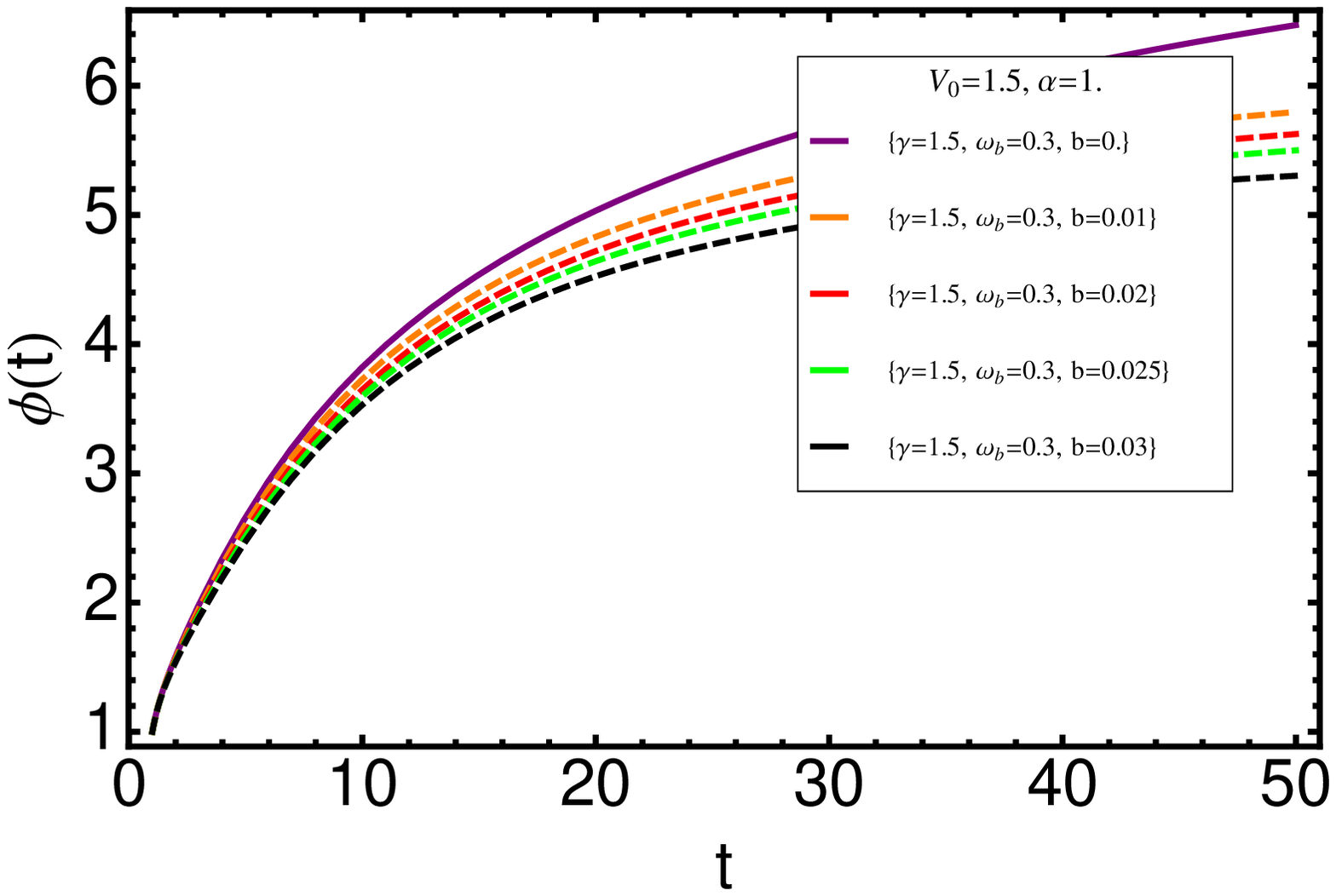} &
\includegraphics[width=50 mm]{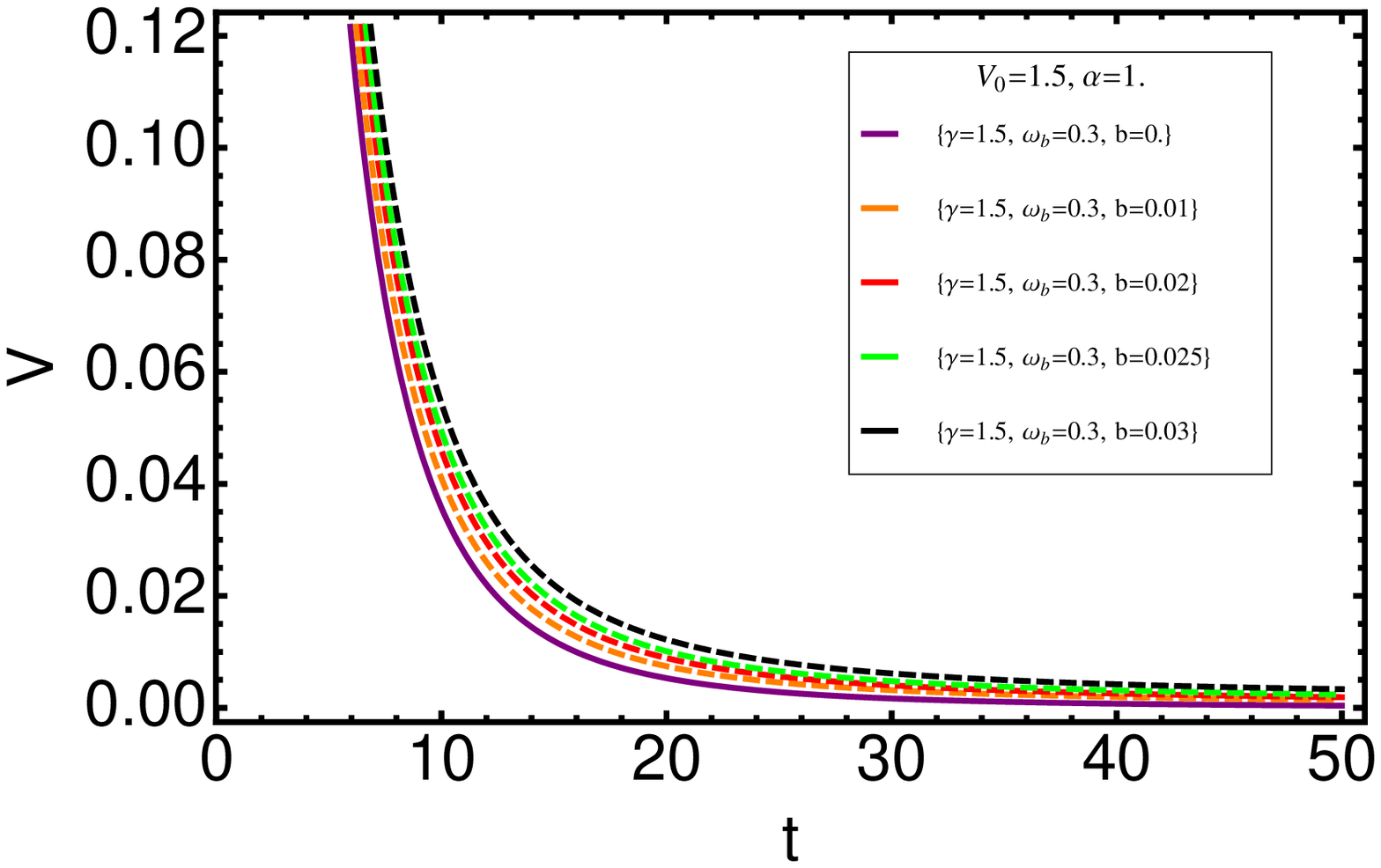}
 \end{array}$
 \end{center}
\caption{Behavior of filed $\phi$ and potential $V$ against $t$ for constant $G$ and $\Lambda$.}
 \label{fig:const3}
\end{figure}

\section{\large{Numerical Results}}
In this section we study behavior of some cosmological parameters in three models by choosing different forms of $\Lambda$.
\subsection{\large{Model 1}}
To describe the dynamics of the universe, we assume that the form of $\Lambda$ is given by Eq. (7). The dynamics of $\beta(t)$ can be obtained from Eq. (\ref{eq:lbeta}) as follow,
\begin{equation}
2\beta\dot{\beta}+6H\beta^{2}+\dot{\rho}_{Q}+\dot{\rho}_{b}-(H+t\dot{H})\rho_{b}e^{[-tH]}=0.
\end{equation}
Our numerical analysis yields to the following results. Plots of the Fig. 4 show behavior of the the Hubble expansion and the deceleration parameters with time. We can see that the the Hubble expansion parameter is decreasing function of time which yields to a constant at the late time. Comparing with the Fig. 1 suggests that consideration of varying $\Lambda$ of the form given by the Eq. (7) decreases value of the Hubble expansion parameter. Also, behavior of the deceleration parameter agree with observational data ($0\geq q\geq-1$) more than the case without interaction. Acceleration to deceleration phase transition also illustrated in the Fig. 4.\\

\begin{figure}[h!]
 \begin{center}$
 \begin{array}{cccc}
\includegraphics[width=50 mm]{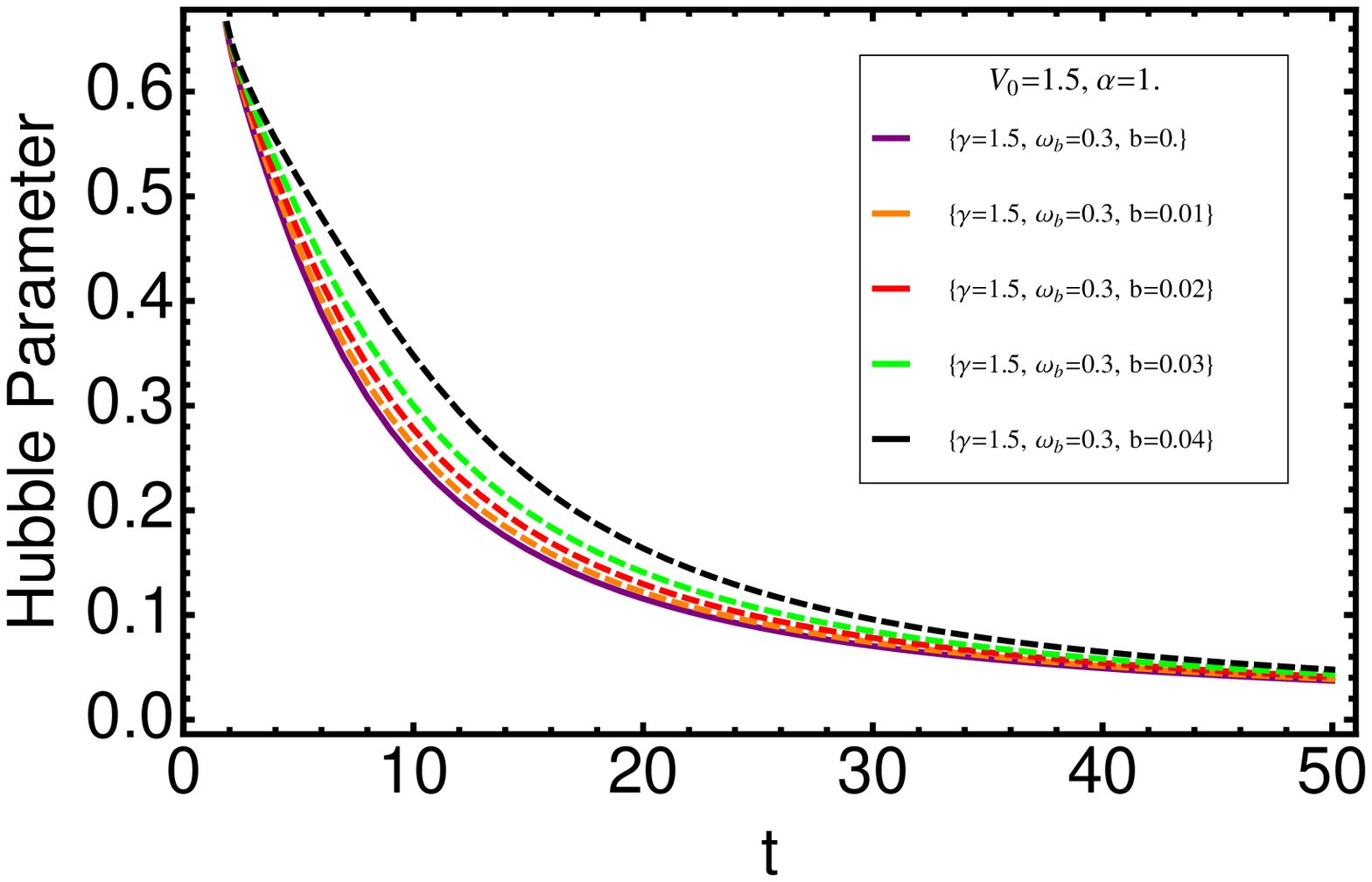} &
\includegraphics[width=50 mm]{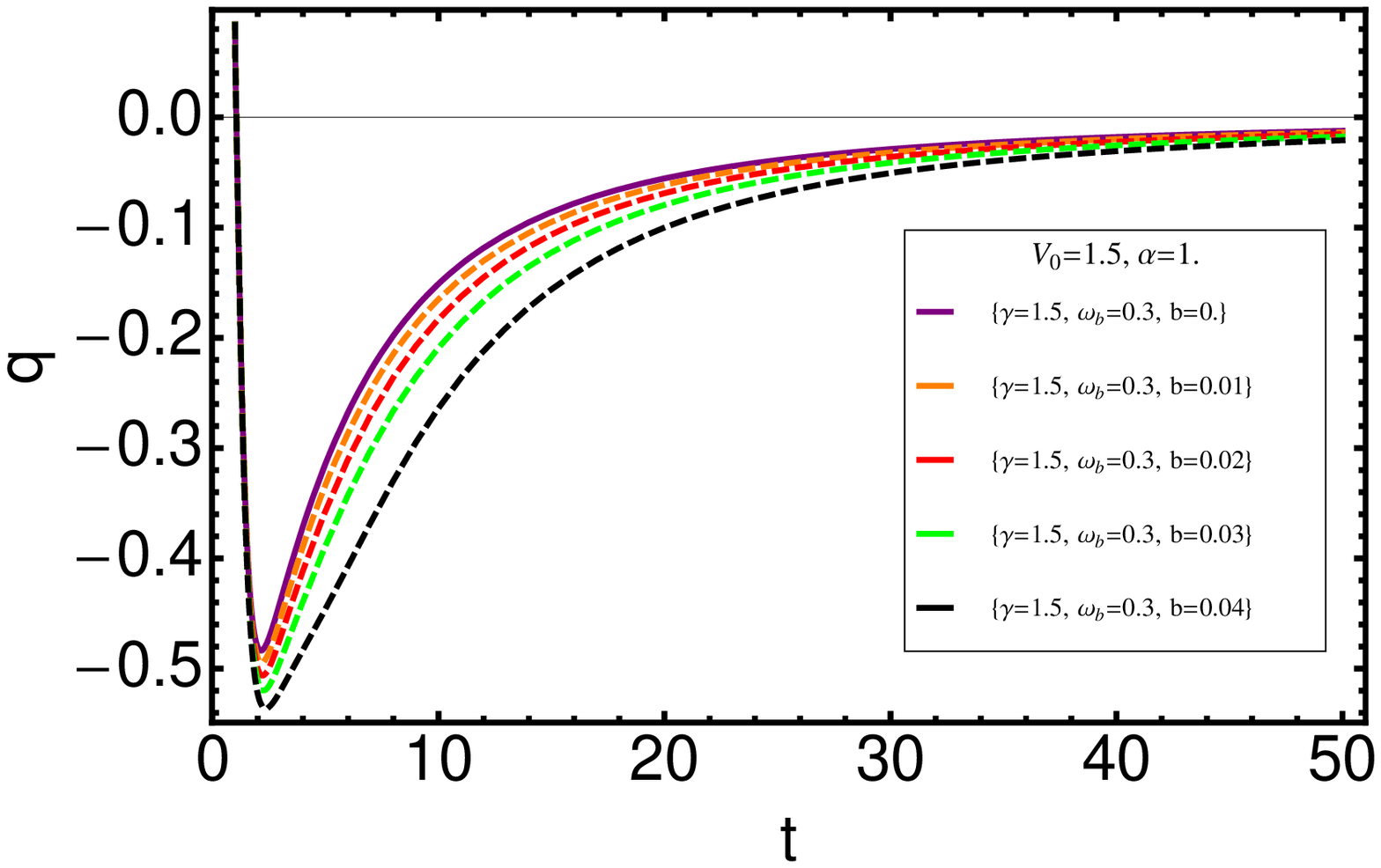}
 \end{array}$
 \end{center}
\caption{Behavior of Hubble parameter $H$ and deceleration parameter $q$ against $t$ for Model 1.}
 \label{fig:1}
\end{figure}

Plots of the Fig. 5 show that value of EoS parameters are within $-1/3\geq\omega\geq-1$. Their value decreased suddenly at the initial time and then grow to reach constant value. This is also agree with observational data.\\
In the Fig. 6 we draw the scalar field $\phi$ and the potential $V$. As expected, the potential vanished at the late time and the scalar field increases by time.\\
We can also obtain the late time behavior of $\beta(t)$ as the following,
\begin{equation}\label{27}
\beta=Ce^{-3Ht},
\end{equation}
where $C$ is an integration constant. In that case we can obtain $C_{s}^{2}=1$, where $C_{s}$ is sound speed. Therefore, we find that the first model is stable, at least at the late time.\\
Our numerical analysis suggest the following densities,
\begin{equation}\label{28}
\rho_{b}=\left(C-\int{\frac{3C^{2}e^{-6(1+H_{0}t)-g(t)}}{t(1+e^{-(1+H_{0}t)})}dt}\right)e^{g(t)},
\end{equation}
where constant $H_{0}$ is current value of the Hubble expansion parameter and,
\begin{equation}\label{29}
g(t)=\int{\frac{(1+H_{0}t)\left(e^{-(1+H_{0}t)}-3(1+\omega)\right)}{t(1+e^{-(1+H_{0}t)})}dt},
\end{equation}
and,
\begin{equation}\label{30}
\rho_{Q}=\left(C-\frac{C^{2}(\frac{3H_{0}}{2t}(3-\omega))^{-\frac{1+3\omega}{4}}e^{-6-\frac{3}{4}H_{0}(3-\omega)t}W(t)}{3H_{0}(3-\omega)(\frac{1+\omega}{2})}\right)t^{-\frac{3}{2}(1+\omega)}e^{-\frac{3}{2}H_{0}(1+\omega)t},
\end{equation}
where,
\begin{equation}\label{31}
W(t)\equiv WhittakerM(\frac{1+3\omega}{4}, \frac{3(1+\omega)}{4}, \frac{3}{2}(3-\omega)H_{0}t).
\end{equation}

\begin{figure}[h!]
 \begin{center}$
 \begin{array}{cccc}
\includegraphics[width=50 mm]{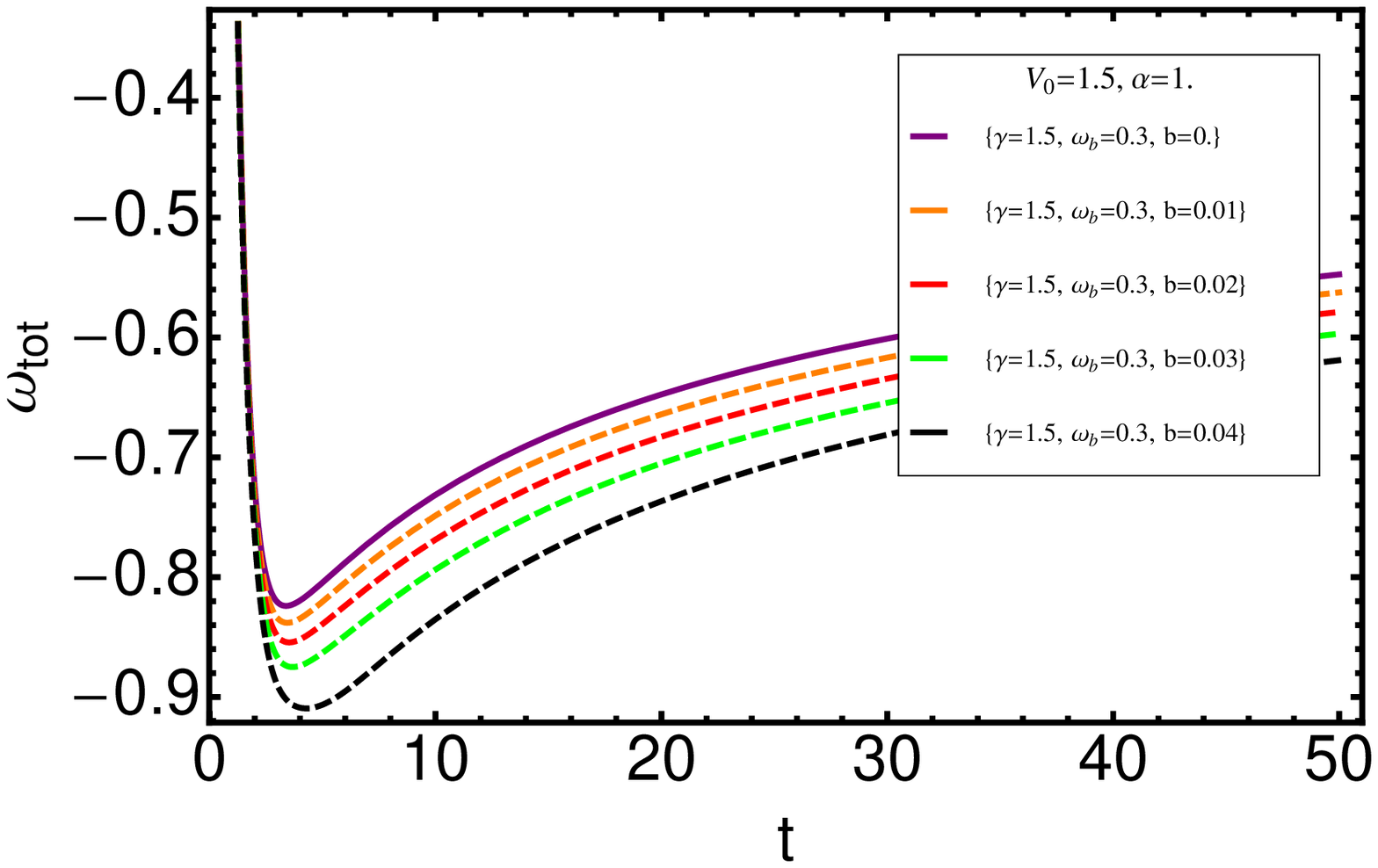} &
\includegraphics[width=50 mm]{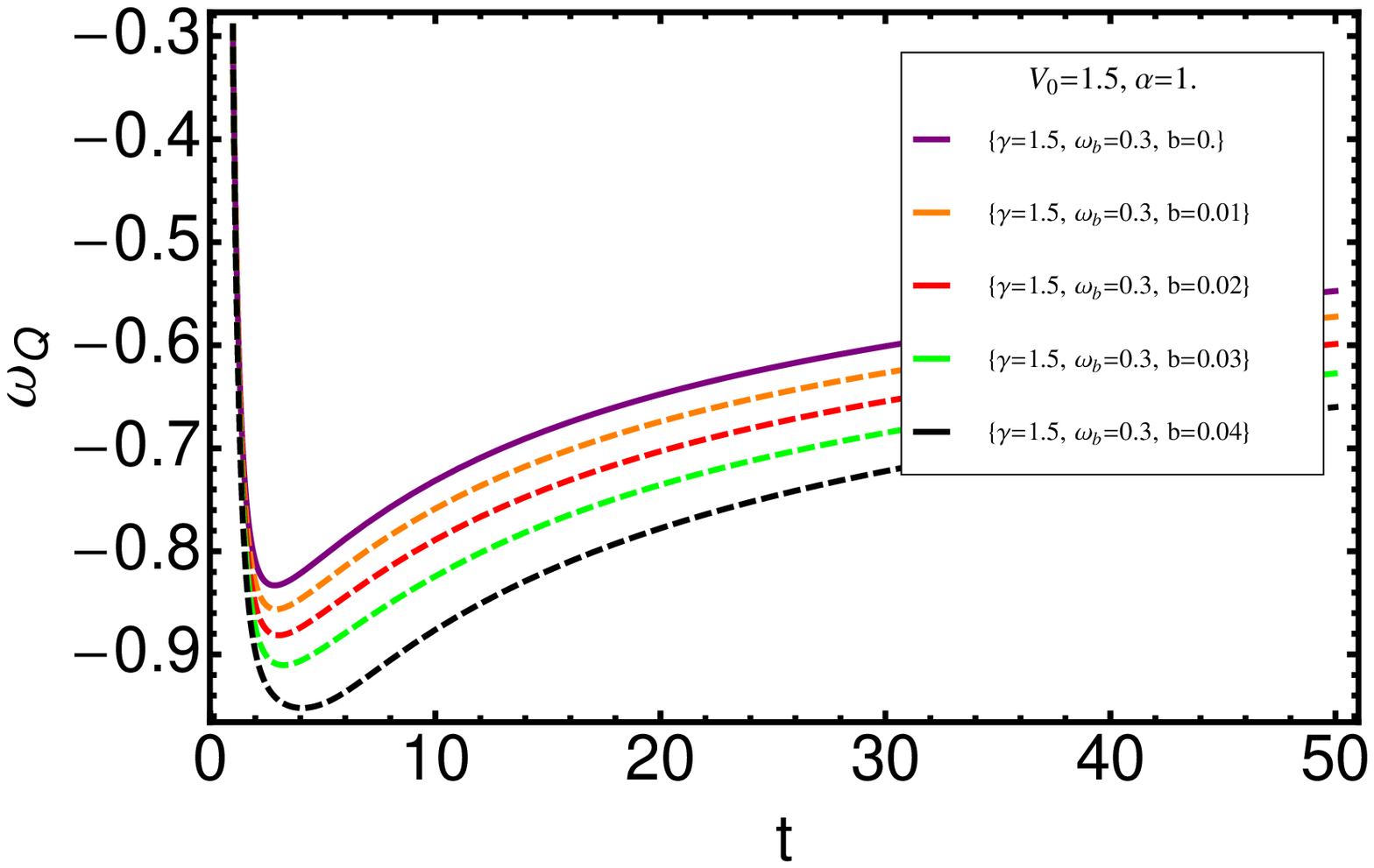}
 \end{array}$
 \end{center}
\caption{Behavior of EoS parameter $\omega_{tot}$ and $\omega_{Q}$ against $t$ for Model 1.}
 \label{fig:2}
\end{figure}

\begin{figure}[h!]
 \begin{center}$
 \begin{array}{cccc}
\includegraphics[width=50 mm]{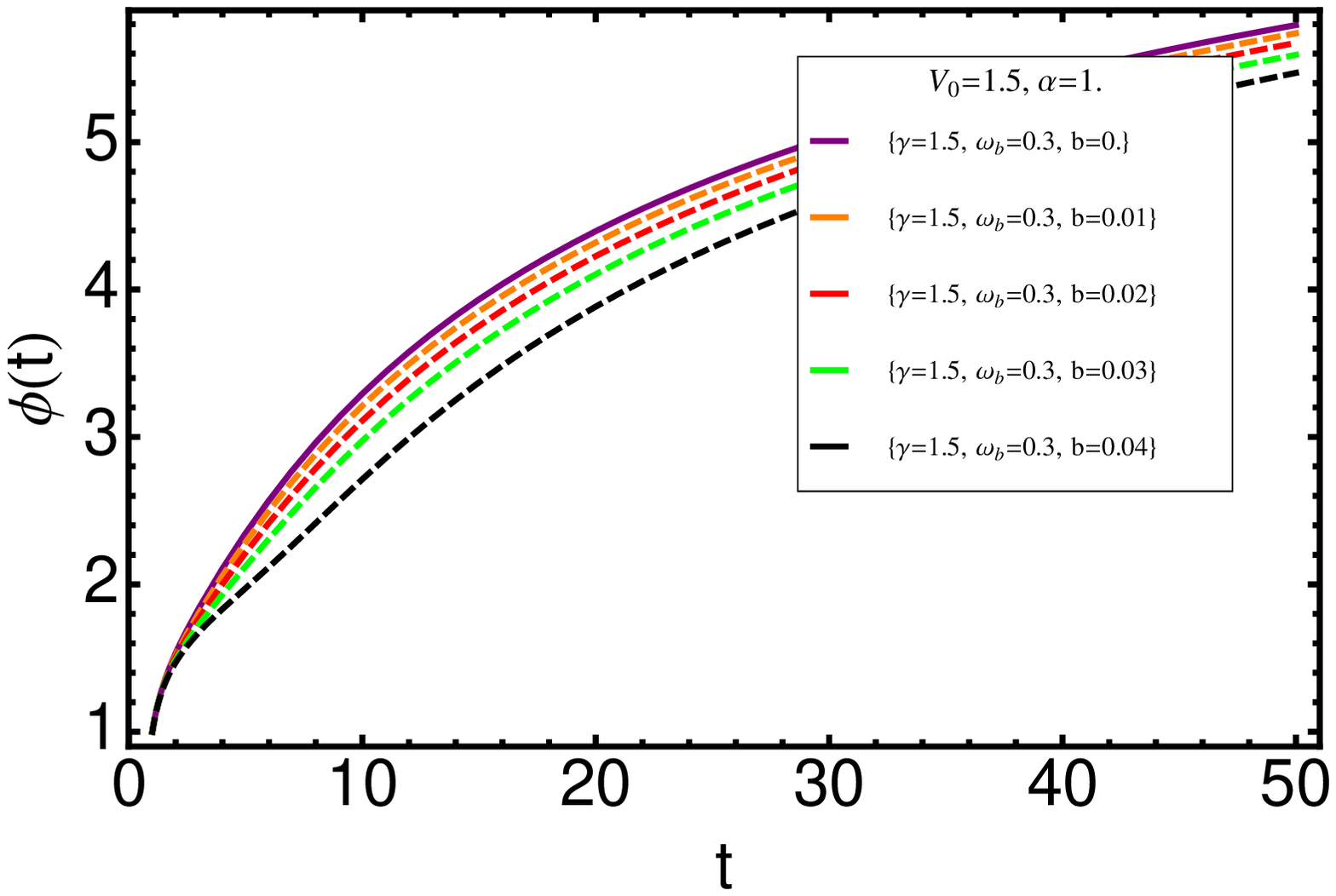} &
\includegraphics[width=50 mm]{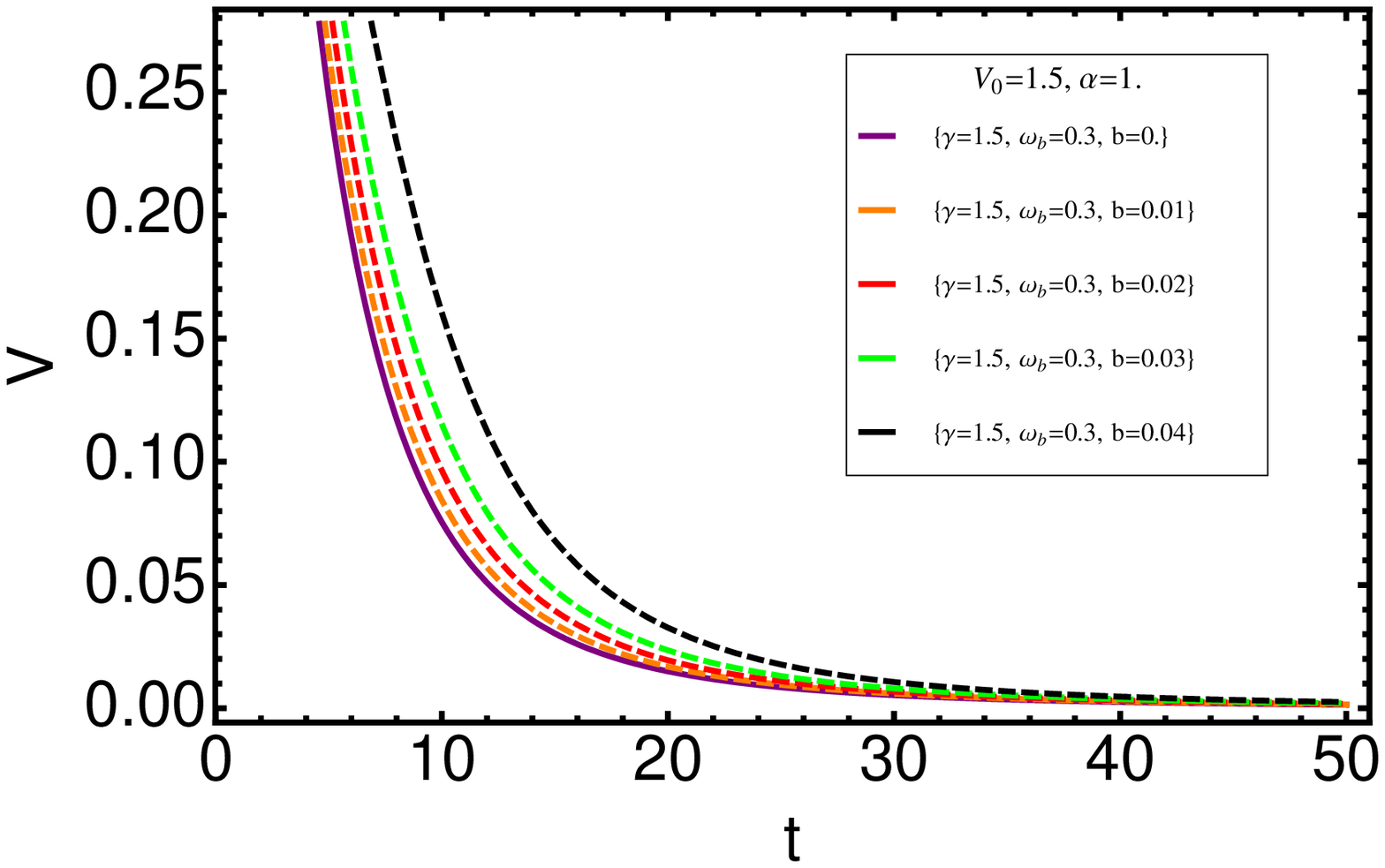} 
 \end{array}$
 \end{center}
\caption{Behavior of filed $\phi$ and potential $V$ against $t$ for Model 1.}
 \label{fig:3}
\end{figure}

\subsection{\large{Model 2}}
In the second model we use the relation (8) for $\Lambda$. The dynamics of $\beta(t)$ can be obtained from Eq. (\ref{eq:lbeta}),
\begin{equation}
2\beta\dot{\beta}+6H\beta^{2}+2H\dot{H}-( ( H+t\dot{H} ) (\rho_{Q}+\rho_{b} ) - \dot{\rho}_{Q}+\dot{\rho}_{b} )e^{[-tH]}=0.
\end{equation}
In the Fig. 7 we can see evolution of the Hubble expansion parameter (left) and deceleration parameter (right). It is  clear that the Hubble parameter is decreasing function of time and yields to a constant value. This is similar to the previous model and lower than the the case without interaction. But, we can see differen behavior of deceleration parameter with previous model. It's value yields to zero at the late time, therefore we can't see agreement with current observational data. Therefore, we left this model and omit presentation of other parameters such as the EoS, the scalar field and the potential, and analyze the next model. However we can obtain behavior of these parameter similar to the previous model.

\begin{figure}[h!]
 \begin{center}$
 \begin{array}{cccc}
\includegraphics[width=50 mm]{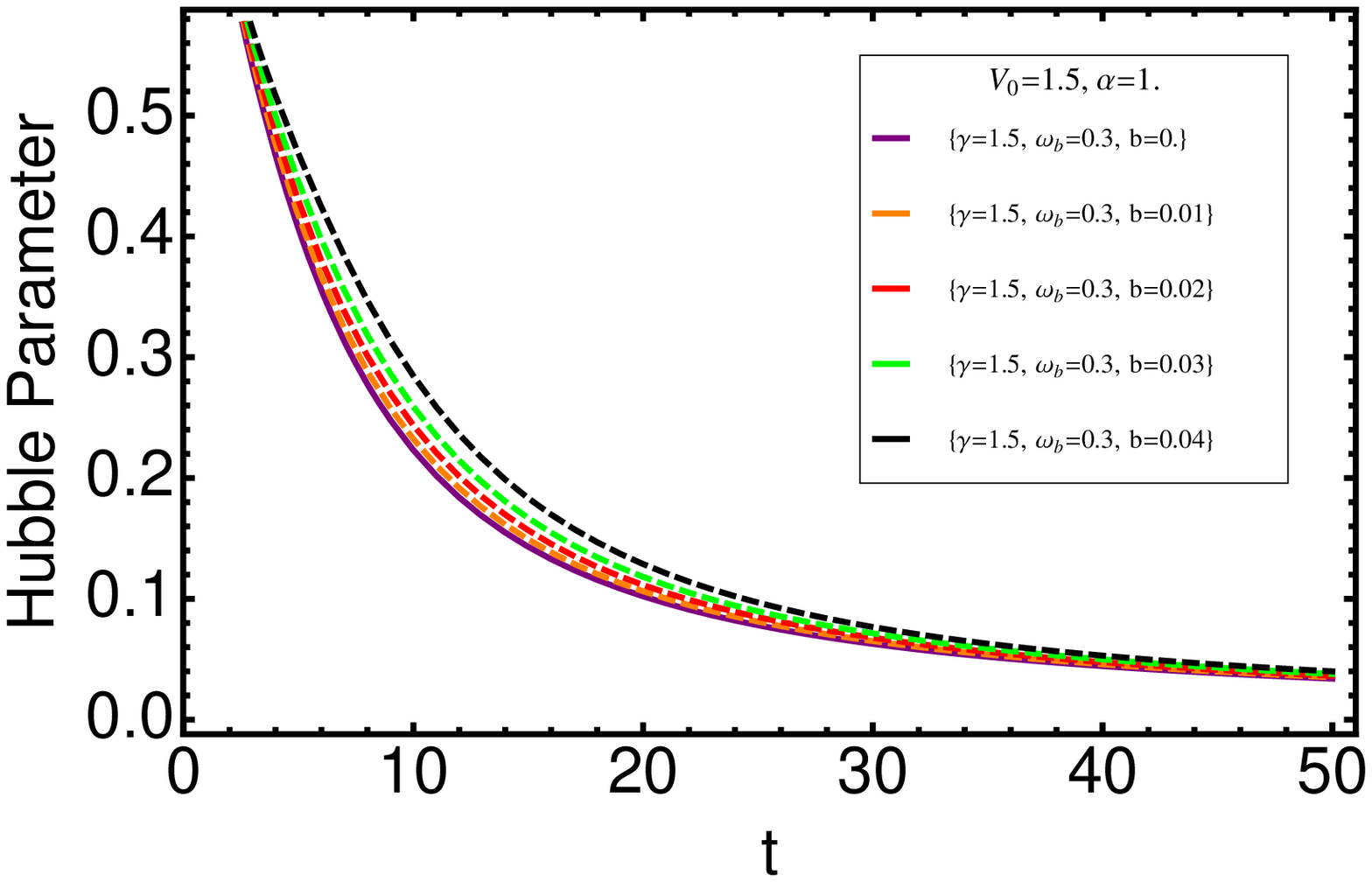} &
\includegraphics[width=50 mm]{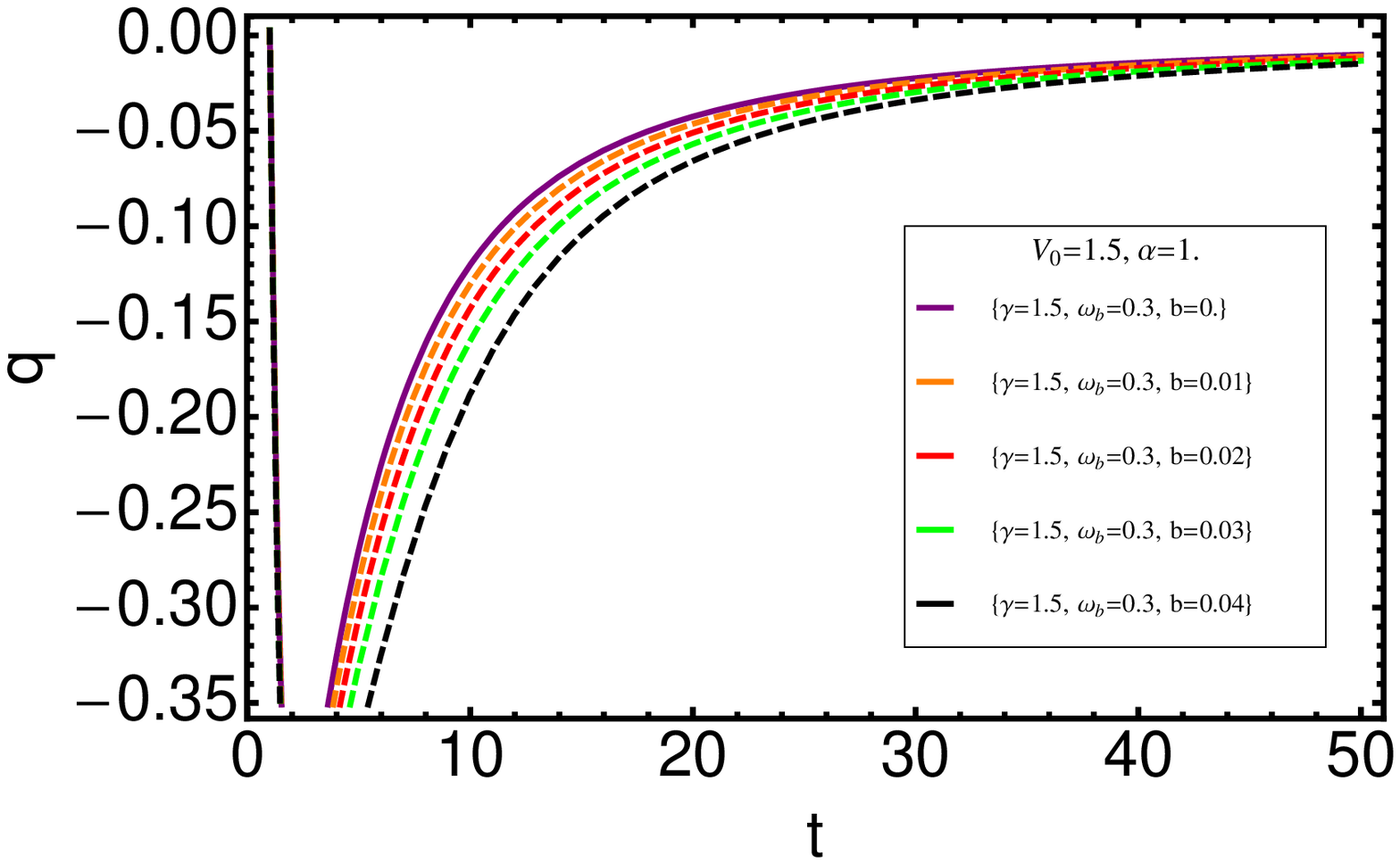}
 \end{array}$
 \end{center}
\caption{Behavior of Hubble parameter $H$ and deceleration parameter $q$ against $t$ for Model 2.}
 \label{fig:4}
\end{figure}

\subsection{\large{Model 3}}
In the third model we use special form of $\Lambda$ which is given by Eq.  (9). Similar to the previous models we can draw plots of $H$ and $q$ (see Fig. 8). We can see similar behavior of the Hubble expansion parameter as previous models but evolution of the deceleration parameter don't show acceleration to deceleration phase transition. Although the value of $q$ obtained within observational data but absence of mentioned transition yields us to left this model like the model 2.

\begin{figure}[h!]
 \begin{center}$
 \begin{array}{cccc}
\includegraphics[width=50 mm]{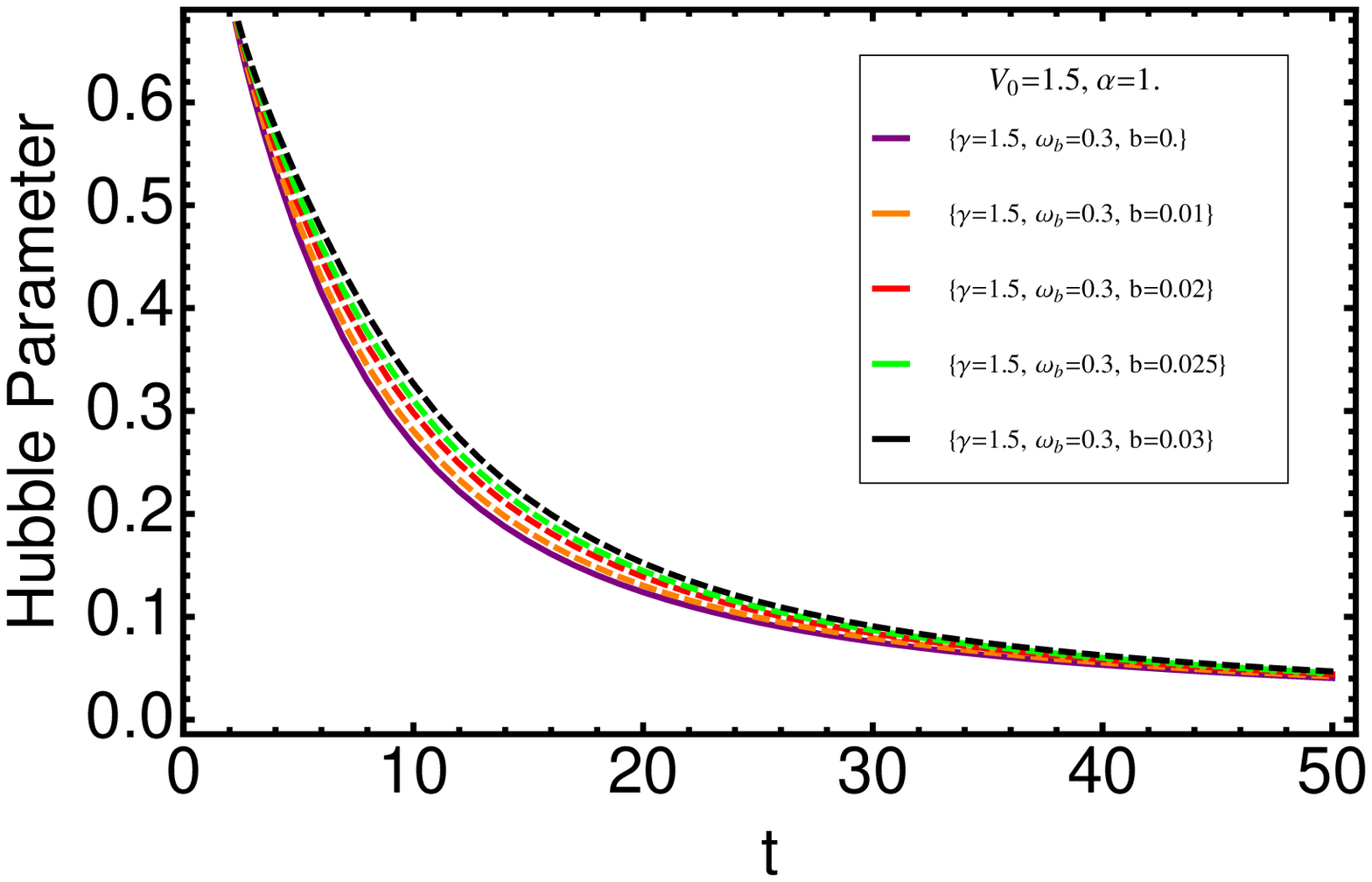} &
\includegraphics[width=50 mm]{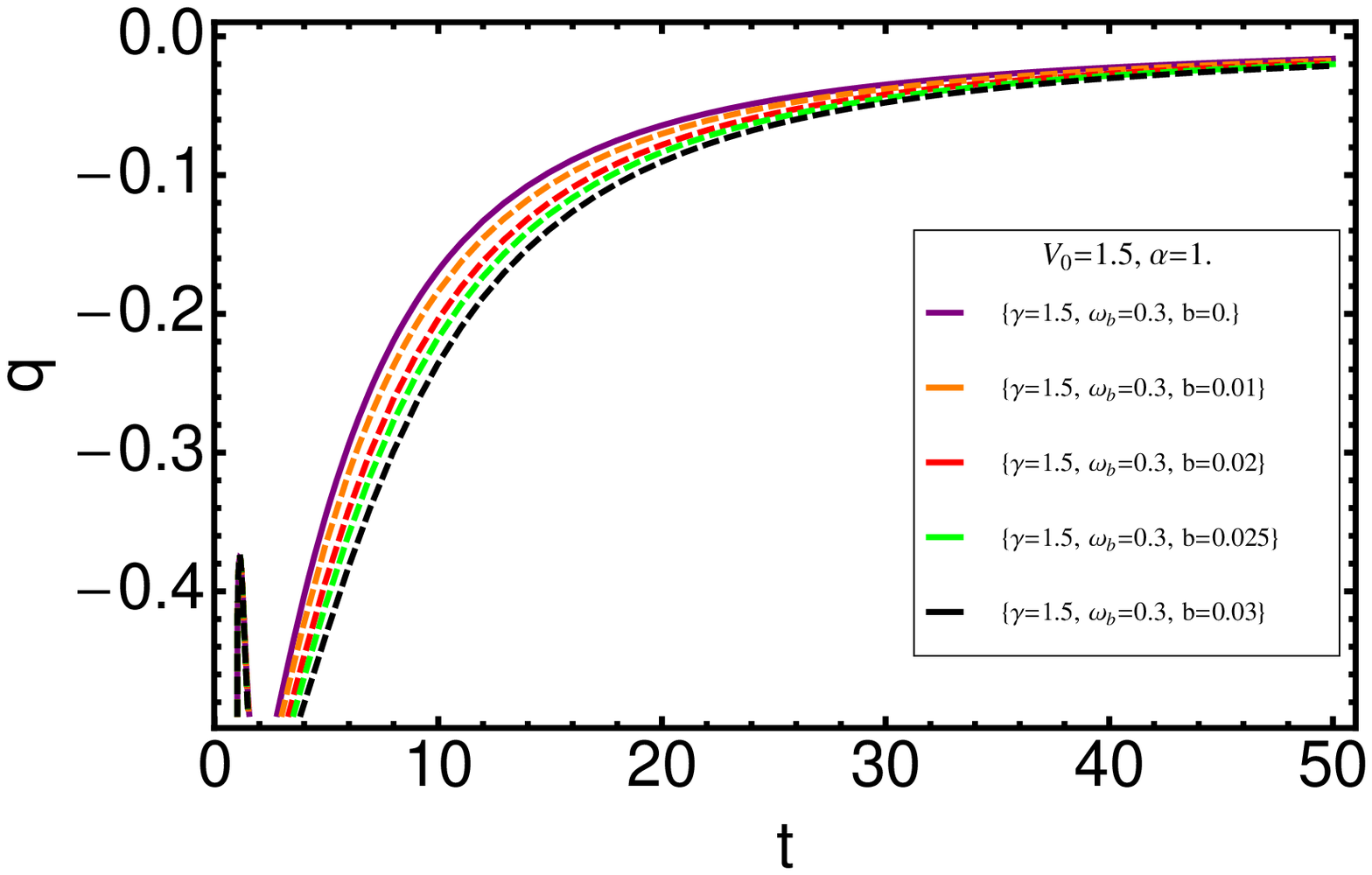}
 \end{array}$
 \end{center}
\caption{Behavior of Hubble parameter $H$ and deceleration parameter $q$ against $t$ for Model 3.}
 \label{fig:7}
\end{figure}

\section{\large{Statefinder diagnostic}}
In the framework of general relativity, one of the properties of dark energy is that it is model-dependent and, in order to choose the best model of dark energy, a sensitive diagnostic tool is needed. The Hubble parameter $H$ and the deceleration parameter $q$ are very important quantities which can describe the geometric properties of the universe. Since $\dot{a}>0$, hence $H>0$ implies an expansion of the Universe. Moreover, $\ddot{a}>0$, which means $q<0$, indicates an accelerated expansion of the universe. Therefore, various dark energy models give $H>0$ and $q<0$, then  they can not provide enough evidence to differentiate the more accurate cosmological observational data and the more general models of dark energy. For this aim, we need  higher order of time derivative of scale factor and geometrical tool. Sahni \emph{et.al} \cite{Sahni} proposed geometrical statefinder diagnostic tool, based on dimensionless parameters $(r, s)$ which are function of scale factor and its higher order time derivatives. These parameters are defined as follow,
\begin{equation}\label{eq:statefinder}
r=\frac{1}{H^{3}}\frac{\dddot{a}}{a} ~~~~~~~~~~~~
s=\frac{r-1}{3(q-\frac{1}{2})}.
\end{equation}
Results of our numerical analysis are presented in the Fig. 9. We can see that the first model has more agreement with observations [37, 38].
\begin{figure}[h!]
 \begin{center}$
 \begin{array}{cccc}
\includegraphics[width=50 mm]{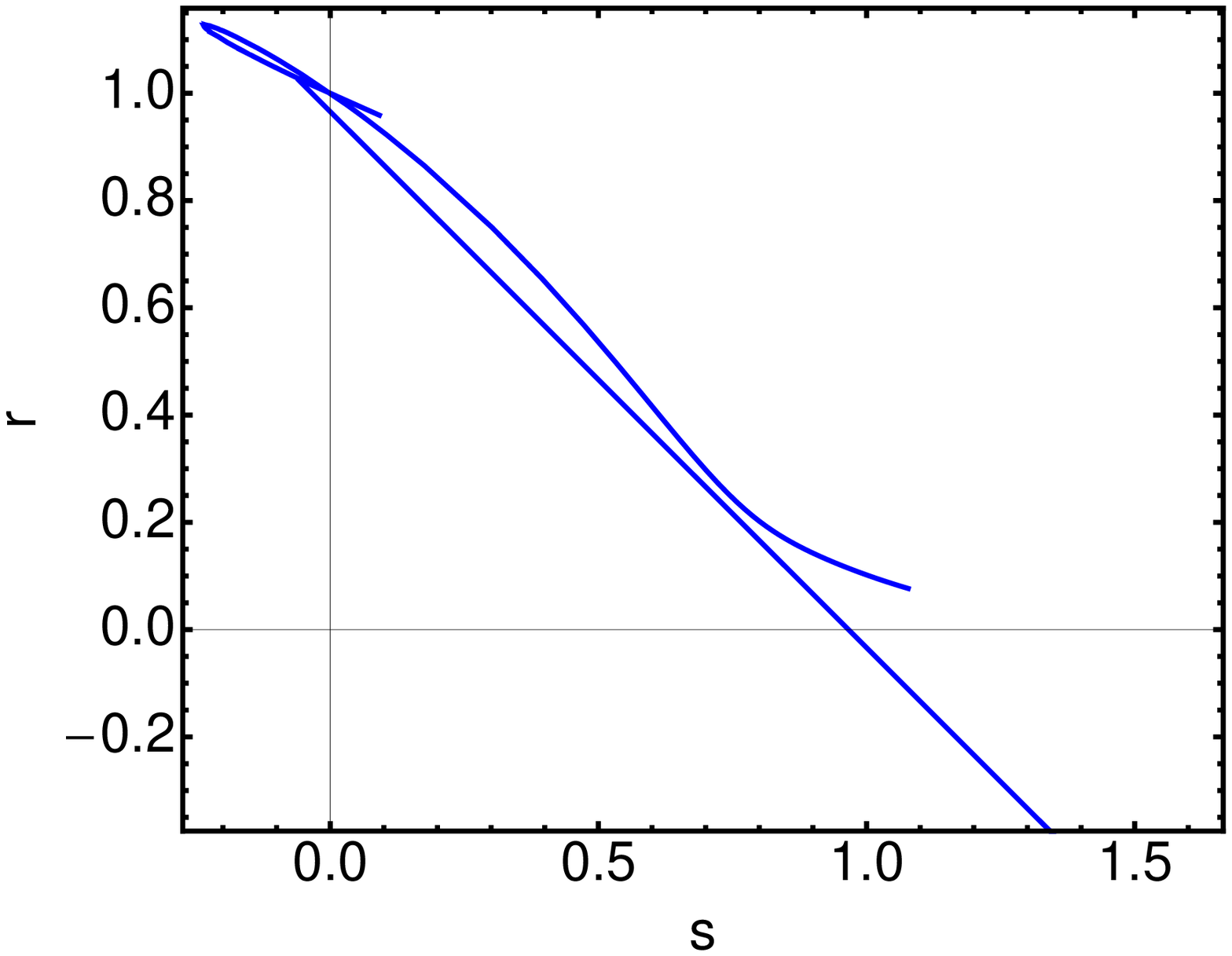} &
\includegraphics[width=50 mm]{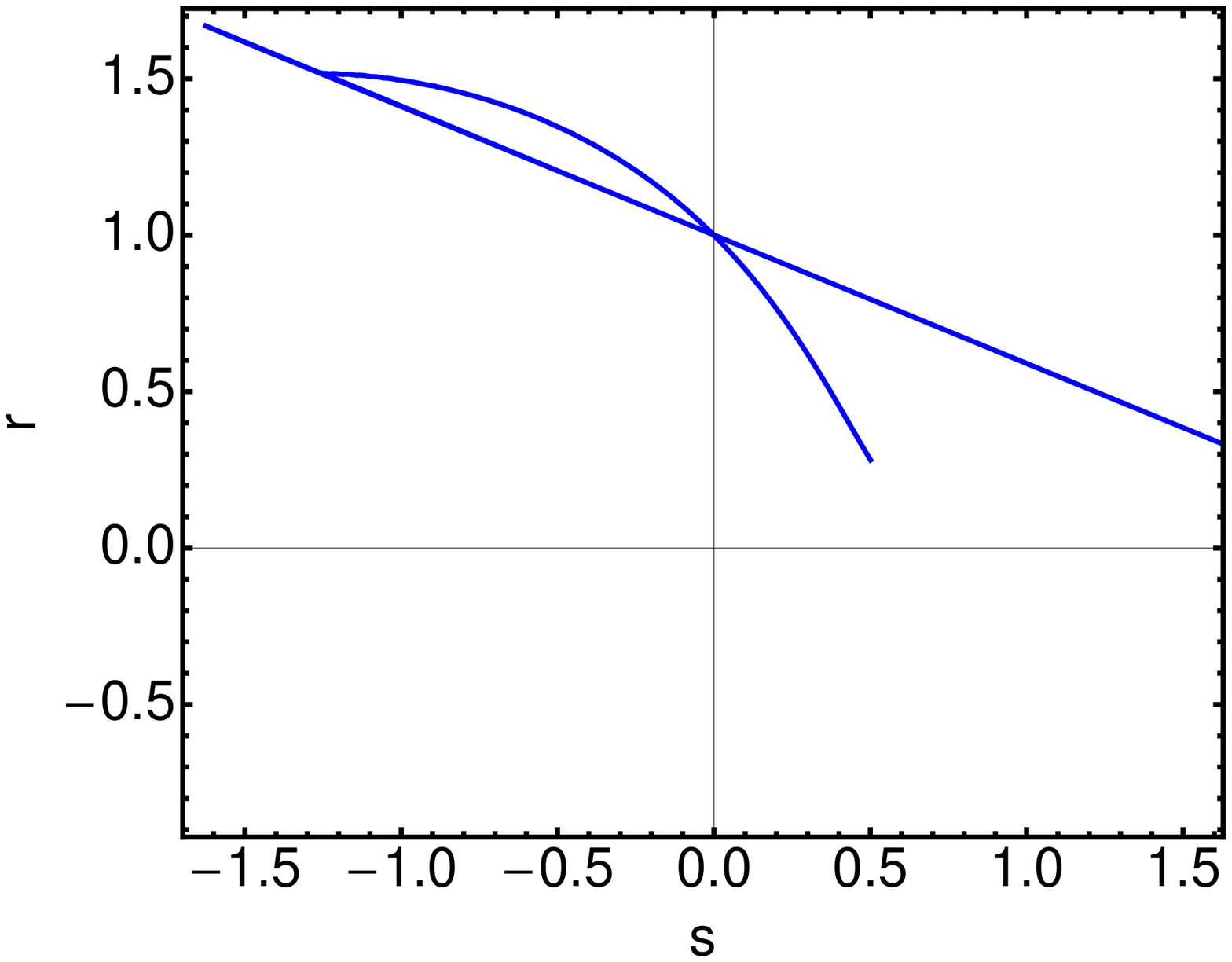}&
\includegraphics[width=60 mm]{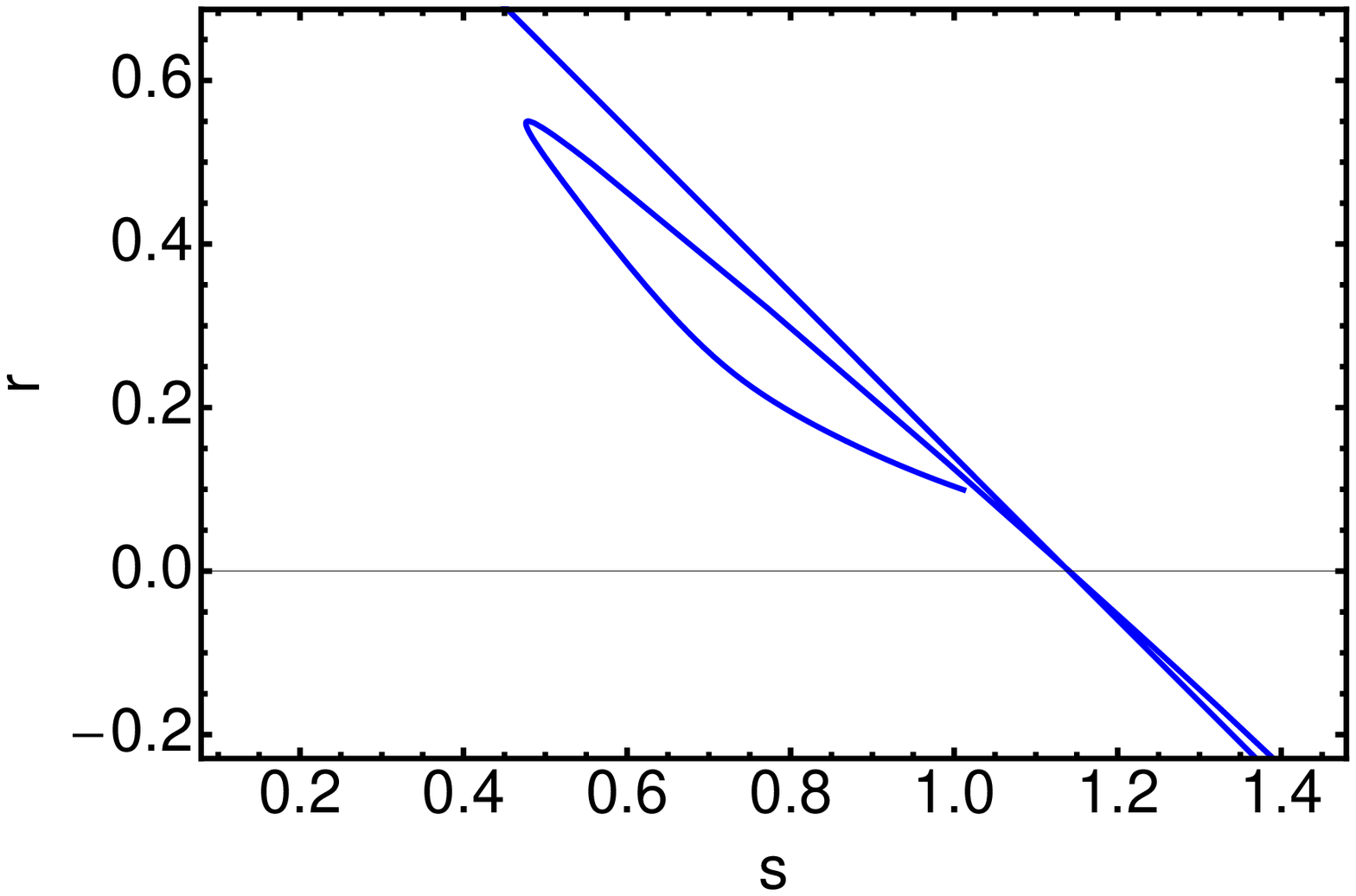}
\end{array}$
 \end{center}
\caption{$r-s$: Model 1 (left), Model 2 (middle), Model 3 (right) .}
 \label{fig:10}
\end{figure}

\section{Discussion}
In this work we considered quintessence cosmology with an effective $\Lambda$-term in Lyra manifold which contains interaction between dark matter and dark energy. We assumed three different forms of variable $\Lambda$. In the first model the variable $\Lambda$ is depend on dark energy density plus dark matter density multiple with exponential function of $(-Ht)$. The second model has variable $\Lambda$ equal to squared Hubble parameter plus total density multiple with exponential function of $(-Ht)$. Finally the variable $\Lambda$ of the third model contains the variable $\Lambda$ of the first model plus inverse of squared time. Therefore late time behavior of the third model expected similar to the first model. Our numerical results show this agreement but acceleration to deceleration phase transition at the early time can't obtained in the third model. Also behavior of the deceleration parameter in the second model is disagree with the observational data. Therefore we conclude that the first model is the best model of this paper which agree with observational data. This point also verified with the statefinder diagnostic tool. We found that the first model is stable at the late time. Comparing our models with the case of constant $\Lambda$ suggests that the first model with varying $\Lambda$ yields to results which are in agreement with observational data more than the case of constant $\Lambda$. Also we found late time behavior of scalar field and obtained dark energy and dark matter densities of the first model by fitting plots.


\begin{thebibliography}{1}
\bibitem{P1}
T. Padmanabhan, Phys. Rept. 380 (2003) 235
\bibitem{P2}
V. Sahni and A.A. Starobinsky, Int. J. Mod. Phys. D 9 (2000) 373
\bibitem{P3}
S. Nobbenhuis, Found. Phys. 36, 613 (2006), [arXiv:gr-qc/0411093]
\bibitem{P4}
C. Armendariz-Picon, V. Mukhanov and P.J. Steinhardt, Phys. Rev.
Lett. 85 (2000) 4438
\bibitem{P5}
A. Sen, "Remarks on tachyon driven cosmology", Phys.Scripta T117 (2005) 70  [arXiv:hep-th/0312153]
\bibitem{P6}
M. C. Bento, O. Bertolami, and A. A. Sen, Phys. Rev. D 66 (2002)
043507
\bibitem{P7}
H. Saadat and  B. Pourhassan, "FRW Bulk Viscous Cosmology with Modified Chaplygin Gas in Flat Space", Astrophysics and Space Science 343 (2013) 783
\bibitem{P8}
A. R. Amani and  B. Pourhassan "Viscous Generalized Chaplygin gas with Arbitrary $\alpha$", Int. J. Theor. Phys. 52 (2013) 1309
\bibitem{P9}
H. Saadat and  B. Pourhassan, "FRW bulk viscous cosmology with modified cosmic Chaplygin gas", Astrophysics and Space Science 344 (2013) 237
\bibitem{P10}
L. Xu, J. Lu, Y. Wang, "Revisiting Generalized Chaplygin
Gas as a Unified Dark Matter and Dark Energy Model", Eur. Phys. J. C
72 (2012) 1883
\bibitem{P11}
B. Pourhassan, "Viscous Modified Cosmic Chaplygin Gas Cosmology" International Journal of Modern Physics D Vol. 22, No. 9 (2013) 1350061 [arXiv:1301.2788 [gr-qc]
\bibitem{P12}
H. Saadat and H. Farahani, "Viscous Chaplygin Gas in Non-flat Universe", Int. J. Theor. Phys. DOI 10.1007/s10773-012-1431-0
\bibitem{P13}
H. Saadat and  B. Pourhassan "Viscous Varying Generalized Chaplygin Gas with Cosmological Constant and Space Curvature", Int. J. Theor. Phys. 52 (2013) 3712
\bibitem{P14}
J. Sadeghi, B. Pourhassan, and Z. Abbaspour Moghaddam, "Interacting Entropy-Corrected Holographic Dark Energy and IR Cut-Off Length", Int. J. Theor. Phys. 53 (2014) 125-135
\bibitem{P15}
J. Sadeghi, B. Pourhassan, M. Khurshudyan, H. Farahani, "Time-Dependent Density of Modified Cosmic Chaplygin Gas with Cosmological Constant in Non-Flat Universe" Int. J. Theor. Phys. 53 (2014) 911
\bibitem{P16}
J. Sadeghi and H. Farahani, "Interaction between viscous varying modified cosmic
Chaplygin gas and Tachyonic fluid", Astrophysics and Space Science 347 (2013) 209
\bibitem{P17}
J. Sadeghi, M. Khurshudyan, H. Farahani, "Phenomenological Varying Modified Chaplygin Gas with Variable $G$ and $\Lambda$: Toy Models for Our Universe", [arXiv:1308.1819 [gr-qc]]
\bibitem{P18}
J. Naji, B. Pourhassan, A. R. Amani "Effect of shear and bulk viscosities on interacting modified Chaplygin gas cosmology", Vol. 23, No. 1 (2013) 1450020
DOI:10.1142/S0218271814500205
\bibitem{P19}
H. Saadat and  B. Pourhassan "Effect of Varying Bulk Viscosity on Generalized Chaplygin Gas", IJTP (2014) [arXiv:1305.6054 [gr-qc]]
\bibitem{P20}
Rong-Jia Yang, "Large-scale structure in superfluid Chaplygin gas cosmology", [arXiv:1312.2416 [gr-qc]]
\bibitem{P21}
P.J.E. Peebles  and B. Ratra, Astrophys. J. Lett., 325 (1988) L17
\bibitem{P22}
M. Khurshudyan, E. Chubaryan and B. Pourhassan, "Interacting Quintessence Models of Dark Energy", Int. J. Theor. Phys. 53 (2014)
\bibitem{P23}
S. Tsujikawa, "Quintessence: A Review", Class. Quant. Grav. 30 (2013) 214003
\bibitem{P24}
V.K. Shchigolev, "Cosmology with an Effective $\Lambda$-Term in Lyra Manifold", Chinese Phys. Lett. 30 (2013) 119801 doi:10.1088/0256-307X/30/11/119801
\bibitem{P25}
R. Chaubey, A.K. Shukla, Int. J. Theor. Phys. 52 (2013)
735
\bibitem{P26}
S.K. Sahu, K. Tapas, Int. J. Theor. Phys. 52 (2013
793
\bibitem{P27}
M. Khurshudyan, E.O Kahya, A. Pasqua, B. Pourhassan, "Higher derivative corrections of f(R) gravity with varying equation of state in the case of variable $G$ and $\Lambda$", [arXiv:1401.6630 [gr-qc]]
\bibitem{P28}
M. Khurshudyan, J. Sadeghi, E. Chubaryan, H. Farahani, "Phenomenologically varying $\Lambda$ and a toy model for the Universe", [arXiv:1401.6270 [gr-qc]]
\bibitem{P29}
M. Khurshudyan, "Phenomenological models of Universe with varying $G$ and $\Lambda$", [arXiv:1311.6898 [gr-qc]]
\bibitem{P30}
J. Sadeghi M. Khurshudyan M. Hakobyan, H. Farahani, "Mutually interacting Tachyon dark energy with variable $G$ and $\Lambda$", [arXiv:1309.7496 [gr-qc]]
\bibitem{P31}
J. Sadeghi, M. Khurshudyan, A. Movsisyan, H. Farahani, "Interacting Ghost Dark Energy Models with Variable $G$ and $\Lambda$", JCAP12(2013)031
\bibitem{P32}
J. Sadeghi, M. Khurshudyan, M. Hakobyan, H. Farahani, "Phenomenological Fluids from Interacting Tachyonic Scalar Fields", [arXiv:1308.5364 [gr-qc]]
\bibitem{P33}
B. Pourhassan, M. Khurshudyan, "Interacting two-component fluid models with varying EoS parameter", [arXiv:1312.1162 [gr-qc]]
\bibitem{P34}
J. Sadeghi, M. Khurshudyan, M. Hakobyan, H. Farahani, "Hubble parameter corrected interactions in cosmology", [arXiv:1310.3421 [gr-qc]]
\bibitem{P35}
J. Sadeghi, M. Khurshudyan, A. Movsisyan, H. Farahani, "Interacting Ghost Dark Energy Models in the Higher Dimensional Cosmology", [arXiv:1401.6649 [gr-qc]]
\bibitem{Sahni}
V. Sahni, T. D. Saini, A. A. Starobinsky, and U. Alam, Statefinder -- a new geometrical diagnostic of dark energy, JETP Lett. \textbf{77}, 201 (2003)
\bibitem{37}
S.D. Katore, and A. Y Shaikh, "Statefinder Diagnostic for Modified Chaplygin Gas
in Plane Symmetric Universe", The African Review of Physics 7 (2012) 0004
\bibitem{38}
K.S. Adhav,  "Statefinder Diagnostic for Variable Modified
Chaplygin Gas in LRS Bianchi Type I Universe", Advances in Mathematical Physics 2012 (2012) 714350
\end{thebibliography}
\end{document}